\documentstyle[12pt,axodraw]{article}
\begin{document}
\def\thebibliography#1{\section*{REFERENCES\markboth
 {REFERENCES}{REFERENCES}}\list
 {[\arabic{enumi}]}{\settowidth\labelwidth{[#1]}\leftmargin\labelwidth
 \advance\leftmargin\labelsep
 \usecounter{enumi}}
 \def\newblock{\hskip .11em plus .33em minus -.07em}
 \sloppy
 \sfcode`\.=1000\relax}
\let\endthebibliography=\endlist

\hoffset = -1truecm
\voffset = -2truecm

\title{\large\bf
Minimum Of QCD Effective Action As Test Of QCD Confinement Parameter}
\author{
{\normalsize \bf A.N.Mitra \thanks{e.mail:
ganmitra@nde.vsnl.net.in} 
}\\
\normalsize  244 Tagore Park, Delhi-110009, India }
\date{}
\maketitle


\begin{abstract}
A new approach to the non-perturbative regime of QCD is proposed by introducing a 
(non-hermitian) field $B$ related to the usual gluon field $A$ by $B_\mu$ = 
$(1+ \sigma \partial_m) A_\mu$  where $m$ goes to zero after differentiation, and 
$\sigma$ is a parameter which `runs' with momentum ($k$). An exact treatment yields 
a structure $[1/k^2 + 2\mu^2/k^4]$ for the gluon propagator, where $\sigma^2$ =
$\pi k^4/9\mu^2\alpha_s$, showing $linear$ confinement in the instantaneous limit. 
This propagator was recently employed to evaluate some basic condensates and their 
temperature dependence (in the cosmological context), which were all reproduced for 
$\mu = 1GeV$ (termed the `confinement scale parameter'), in association with the 
QCD scale parameter $\Lambda_{qcd}= 200 MeV$ [hep-ph/0109278]. This paper seeks to      
provide a formal basis for the ratio $\Lambda_{qcd} / \mu$  by employing the 
minimality condition for the $integrated$ effective action $\Gamma$, up to the 
2-loop level, using the Cornwall-Jackiw formalism for composite operators. 
To that end the mass function $m(p)$, determined via the Schwinger-Dyson equation 
(as a zero of the functional derivative of $\Gamma$ w.r.t $S_F'$), acts as a feeder,
and the stationarity condition on  $\Gamma$ as function of $\mu$ and $\alpha_s(\mu)$ 
gives the ratio $\Lambda /\mu$ = $0.246$, in fair accord with the value $0.20$  
given above. Inclusion of a two-loop $\Gamma$ is crucial for the agreement.  \\    
Keywords: confinement scale, QCD effective action, minimality condition, $B$-field.
\end{abstract}

\section{Introduction}

\setcounter{equation}{0}
\renewcommand{\theequation}{1.\arabic{equation}}

While the high momentum regime of QCD characterizing asymptotic freedom is amenable to 
precise calculations by pQCD techniques via RG equations, the opposite regime of low momentum
confinement is unfortunately not so easily accessible, despite notable efforts by stalwarts 
[1-7] almost since the birth of QCD [8], employing ideas dating further back [9,10].      
The central issue of chiral symmetry breaking is essentially one first suggested by
NJL [9] on the lines of BCS-type superconductivity showing up through $q{\bar q}$ pair
condensation. The latter is supposed to come about by the attraction between quarks and
antiquarks mediated by single gluon exchange if the running coupling constant $\alpha_s$
exceeds a critical value [1,2], or alternatively by the instanton mechanism [3]. The closely
related issue of confining forces being the cause of $\gamma_5$-breaking [4], is probably
more academic [5] than the fact that both these features are brought about by a common 
mechanism, viz., the occurrence of large enough $\alpha_s$ in single gluon exchange [1,2]. 
\par
	In this paper we shall be concerned with the $confinement$ aspect of gluon
exchange forces, more explicitly with $linear$ confinement (which among other things
is known to be a key mechanism for the understanding of heavy quark spectroscopy).     
The incentive for a non-perturbative approach of this kind stems from a recent exercise 
[11] undertaken to incorporate linear confinement in the operational mode of QCD, namely 
a $k^{-4}$ behaviour for the gluon propagator as a means of calculating afresh the principal 
QCD condensates, together with their $T$-dependence, in view of a lack of consensus in the 
results from usual non-perturbative approaches to these quantities in the cosmological context [11].  
The results of this exercise were encouraging enough to warrant a more microscopic study of 
the anatomy of the confinement proposal [11] within the QCD framework, with a view to 
put it on a firmer foundation, which is the main purpose of this paper.
\par
	To recall the essential features of the proposal in ref.[11], but in a slightly
changed notation, one defines a new (non-hermian) field $B_\mu$, $B_\mu^\dag$ related to 
the actual gluon field $A_\mu$ of QCD by: 
\begin{equation}\label{1.1}
B_\mu =(1+ \sigma \partial_m) A_\mu; \quad B_\mu^\dag = A_\mu (1- \sigma\partial_m)
\end{equation}  
wherein the derivative is w.r.t. a small gluon mass $m$ that goes to zero after  
differentiation. The parameter $\sigma$, by the very nature of its appearance, is 
suggestive of  a (low energy) confinement scale which must be intimately related 
to the more fundamental QCD parameter $\Lambda_{qcd}$ that is governed by RG theory,
although to show the connection between the two is a non-trivial task. It will be 
shown in this paper (see Appendix A) that the full gluon propagator derivable
from (1.1) has the following form in momentum space:  
\begin{equation}\label{1.2}
\Delta_F (k^2) = [\frac{1}{k^2} + \frac{2\mu^2}{k^4}] 
\end{equation}  
where $\mu$ is related to $\sigma$ as:
\begin{equation}\label{1.3}
\sigma^2 = \frac{\pi k^4}{9 \mu^2 \alpha_s}
\end{equation} 
The form (1.2) was employed in ref.[11] where  a value of $\mu = 1 GeV$ 
(reminiscent of the universal Regge slope), yielded results consistent with QCD-SR [12]
and chiral perturbation theory ($\chi PT$) [13] for $T=0$, while keeping $\Lambda_{qcd}$ 
fixed at the more or less standard value of $200 MeV$ [14]. We shall now seek to find 
a formal link of the $\mu$ parameter with the more fundamental QCD scale parameter 
$\Lambda_{qcd}$, using the principle of minimal effective action. Before describing 
this method, we note that the piece $2\mu^2 k^{-4}$ in (1.2) corresponds to $linear$ 
confinement. This fact may be checked  by a Fourier transform of (1.2) to the 
$r$-representation  in the $instantaneous$ limit ($t=0$) which gives for 
the sum of the o.g.e. and confinement propagators in the $r$-representation, a potential 
$V(r)$ obtained as a limit of $m=0$ through the following steps: 
\begin{eqnarray}\label{1.4}
V(r) &=& \int d^3 {\bf k} dk_0 \exp{i({\bf k}.{\bf r}- k_0 t)} [ \frac{1}{{\bf k}^2 + m^2}+ 
\frac{2\mu^2}{({\bf k}^2 + m^2)^2}]  \\  \nonumber      
     &=& \delta (t) \int d^3{\bf k}\exp{i{\bf k}.{\bf r}}[1-(\mu \partial_m)^2] \frac{1}{{\bf k}^2 +m^2}\\ \nonumber    
     &=& 4 \pi^2 \delta(t) [1-(\mu \partial_m)^2] \frac{e^{-m r}}{r} 
      \Rightarrow 4 \pi^2 \delta(t) [ \frac{1}{r} - \mu^2 r ]
\end{eqnarray}   
As a consistency check, the o.g.e. and confinement terms come with $opposite$ signs in $r$-space, while 
coming with the $same$ sign in $k$-space. (This derivation may be compared with that of Gromes [15]).
Note that the linear confinement is a 3D concept (by virtue of the instantaneous limit), while the 4D
coordinate form of the $k^{-4}$ behaviour is logarithmic, as can be easily checked [4,7].    
\par    
	To relate the $\mu$ parameter to the QCD scale parameter $\Lambda_{qcd}$ 
governed by RG theory, we shall employ the principle  of minimality of the QCD 
effective action $\Gamma$ via the Cornwall-Jackiw-Tomboulis (CJT) [16] formalism for composite 
operators, as given in Miransky [17]. Now the minimality condition on $\Gamma (G,\phi)$ can be 
treated at $two$ distinct levels. At the first (usual) level, setting the functional derivative 
of $\Gamma$  w.r.t. the quark's Green's function $G$ to zero, gives rise to the 
Schwinger-Dyson equation (SDE) for the mass function $m(p)$. At a second (less conventional) 
level, the $integrated$ effective action $\Gamma$ w.r.t. the loop momenta may be regarded as 
an ordinary function of its input parameters $\mu, \Lambda_{qcd}$, so that its  minimality 
w.r.t. $\mu$ should give its desired connection with $\Lambda_{qcd}$. For
this (second level) determination, the (first level) SDE acts as the feeder in
which the mass function $m(p)$ plays a central role, with $m(0)$ (the $constituent$ mass [4]) 
expressed in terms of $\mu$ and $\alpha_s(\mu)$. It should suffice to  go up to two-loop 
irreducible diagrams for a realistic determination. 
\par
	In Sect.2, we sketch the CJT [16] formalism as given in Miransky [17],
and set up the effective action up to the two-loop level in a notation closely 
following ref. [17], and obtain the mass function $m(p)$ as a solution of the SDE.
In Sects.3 and 4, we calculate the integrated effective action for the one- and 
two-loop contributions respectively, using the results of Sect 2 for $m(p)$, as
well as the  dimensional regularization method of t'Hooft and Veltman [18]. 
Sect 5 gives our principal result, viz.,$\Lambda_{qcd} = 0.246 \mu$,  from the 
minimality of the integrated action, (to be compared with the value $0.20$ obtained
from the principal condensates [11]), and concludes with a discussion of this
method vis-a-vis other contemporary non-perturbative approaches.  
       
\section{Effective Action For Composite Operators}

\setcounter{equation}{0}
\renewcommand{\theequation}{2.\arabic{equation}}

We first summarize the results of the basic CJT [16] formalism for the effective 
action for composite operators, as enunciated by Miransky [17] (Chapter 8), with a view 
to adapting it to QCD, by incorporating the structure (1.2) of the full gluon propagator
(including confinement). To fix the ideas, the effective action $\Gamma$ is a 
functional of both the vacuum average $\phi_c(x) = <0|\phi(x)|0>$, and the propagator 
$G(x,y) = i<0|T\phi(x)\phi(y)|>$ corresponding to $\phi$ (a generic name for
the collection of fields in a given Lagrangian). The minimum for effective action
is expressed by the zeros of the functional derivatives
\begin{equation}\label{2.1}
\frac{\delta \Gamma}{\delta \phi_c(x)} = 0; \quad \frac{\delta \Gamma}{\delta G(x,y)} =0      
\end{equation}
Now the functional $\Gamma$ admits a loop expansion (up to 2-loops) of the form [16,17]
\begin{equation}\label{2.2}
\Gamma(\phi_c, G) = S(\phi_c) + \frac{i}{2} Tr \ln G^{-1} + \frac{i}{2} Tr [ D^{-1} G] \\
+ \Gamma_2(\phi_c, G) + C
\end{equation}
$S(\phi_c)$ being the classical action, $D$ the lowest order propagator, and $\Gamma_2$
the effective action in the 2-loop order. The $Tr$ in each term stands for the 
summation over all the internal variables (spin, polarization), including integration
over momenta. For translationally invariant solutions, with $\phi_c$  constant, this
parameter may henceforth be dropped. Further, the `effective potential' $V (G)$  is 
merely the reduced effective action after  taking out the 4D $\delta$-function
(i.e. the 4D volume element) from the latter [17].   
\par
	To adapt $\Gamma$ to QCD, which has two distinct fields, 
quarks (with full propagator $S_F'$), and gluons (with full propagator $\Delta$), we 
may use the QCD version of the QED form given by eq.(8.57) of ref [17] in an obvious matrix 
notation, namely,
\begin{equation}\label{2.3}
\Gamma(S_F', \Delta) = i Tr [\ln {\frac{S_F}{S_F'}} + \frac{S_F'}{S_F} - \frac{1}{2} 
\ln{\frac{D}{\Delta}}-\frac{1}{2} \frac{\Delta}{D}] + \Gamma_2 (S_F', \Delta) + C
\end{equation}
where $S_F$ and $D$ are the unperturbed quark and gluon propagators respectively,
and we have normalized the arguments in the respective logarithms to their unperturbed values.
These functions, in momentum space, are defined as: 
\begin{equation}\label{2.4}
i S_F' (p) = \frac {m(p)- i\gamma.p}{m^2(p)+ p^2};\quad iS_F(p) = \frac{1}{i\gamma.p}; 
\end{equation}
\begin{equation}\label{2.5}
i\Delta_{\mu\nu}(k) = \frac{\delta_{\mu\nu}- k_\mu k_\nu / k^2}{(k^2 + m^2)}[1+2\mu^2/(k^2 + m^2)]; \quad 
   iD_{\mu\nu}(k)   =  \frac{\delta_{\mu\nu}- k_\mu k_\nu / k^2}{k^2 + m^2}
\end{equation}
   
Here we have taken the Landau gauge, for which the $A(p)$ function is unity [17, 19], so that 
the $B(p)$ function may be directly read as the mass function $m(p)$. The small 
quantity $m$ in the gluon propagator tends to zero at the end, while the current 
mass of the (light) quark has been ignored. Then substituting  (2.4-5) in (2.3), and 

evaluating the traces a la ref [17],  eq.(2.3) may be written as an `effective potential'
$V$ as a sum of the one-loop ($V_1$) and two-loop ($V_2$) contributions as momentum integrals
in Euclidean space. The one-loop contribution is
\begin{eqnarray}\label{2.6}
V_1(S_F',\Delta) &=& \int \frac{2d^4 p}{(2\pi)^4} [- \ln{\frac{1 + m^2(p)}{p^2}} - 2 \frac{p^2}{m^2(p) +p^2}  
\\  \nonumber
                 & & - \ln{(1+\frac{2\mu^2}{p^2 + m^2})} + (1+ \frac{2\mu^2}{p^2 +m^2})]
\end{eqnarray}
            
The two-loop contribution, in which gluon line is inserted within a quark loop [c.f. fig 8.5 (a) of
ref [17]  in coordinate space], is (eq.(8.59) of ref.[17])
$$
\Gamma_2 = \frac{ g_s^2 F_1.F_2}{2} \int d^4xd^4y tr[ S_F'(x,y)\gamma_\mu S_F'(y,x)\gamma_\nu \Delta_{\mu\nu}]
$$
noting that a corresponding diagram with a gluon line joining two separate quark loops
does not contribute [17]. The color factor $F_1.F_2$ has the value $(-4/3)$. Written out in momentum 
space in the same notation and normalization as above, the contribution to the two-loop
effective potential becomes
\begin{equation}\label{2.7}
V_2 (S_F', \Delta)  = \frac{g_s^2 F_1.F_2.}{2} \int \int \frac{d^4 p d^4 k}{(2\pi)^8} 
 Tr[ S_F'(p) \gamma_\mu S_F'(p-k) \gamma_\nu \Delta_{\mu\nu}]
\end{equation}
where the momentum space functions are given by (2.4-5). The evaluation of (2.7) for $V_2$ is
described in Sect 3. In the remainder of this Section we show the evaluation of $V_1$
after summarizing and refining the results of [11] on $m(p)$ via the SDE. 

\subsection{SDE And The Structure of $m(p)$}      

From ref [11], the full SDE, including the confining interaction of Eq. (1.2), is:
\begin{equation}\label{2.8}
m(p) = -3g_s^2 F_1.F_2 [1 - \mu^2 \partial_m^2] \int \frac{-i d^4 k}{(2\pi)^4} \frac{m(p-k)}{(m^2+k^2)[m^2(p-k)+(p-k)^2]}
\end{equation}
where $g_s^2 = 4\pi \alpha_s$, $F_1.F_2 = -4/3$ is the color
Casimir; the Landau gauge has been employed [19], and $m=0$ after
differentiation. Our defense of the Landau gauge is essentially
one of practical expediency, since this gauge usually offers the
safest and quickest route to a gauge invariant result, even
without a detailed gauge check, for there has been no conscious
violation of this requirement at any stage in the input
assumptions. For an approximate solution of this equation, we adopt
the following strategy.  As a first step, we
replace the mass function inside the integral by $m(p)$.  Then the
method of Feynman for combining denominators with an auxiliary variable $u$ 
and a subsequent translation  $k \rightarrow k+ p u$ yields an integral
which can be treated [11]  by the method of dimensional regularization (DR) [18] 
after the generalization $4 \rightarrow n$. The result before integration is
\begin{equation}\label{2.9}
m(p) = +4 g_s^2 \int \frac{d^n k}{(2 \pi)^n}\int_0^1 du \frac{m(p)[1 - \mu^2 \partial_m^2]}
     {[m^2(p)u + m^2 (1-u) + p^2 u(1-u)]^2}
\end{equation}          
the integral on the RHS of Eq. (2.9) may be carried out by using the following formulae:
\begin{equation}\label{2.10}
\int \frac{d^n p}{(2\pi)^n} \frac{1}{(a p^2 + b)^\alpha}= \frac{(b\pi/a)^{n/2}}{(2\pi)^n b^\alpha}
B(n/2, \alpha - n/2);
\end{equation}
\begin{equation}\label{2.11}
\int \frac{d^n p}{(2 \pi)^n} ln(a p^2 + b) = - \frac{(b\pi/a)^{n/2}}{(2 \pi)^n} \Gamma(-n/2)
\end{equation}   
The resulting function of $n$ has a pole at $n=4$ which should be subtracted a la ref [18]
by putting $n-4=\epsilon$ and expanding in powers of $\epsilon$. The final result
after the operation of $\mu^2 \partial_m^2$ and striking out $m(p)$ from both sides, is
\begin{eqnarray}\label{2.12}
\frac{\pi}{\alpha_s(\mu)} &=& \int_0^1 du[\mu^2 / \Omega(u) -\gamma - ln[\Omega(u) / \mu^2] \\  \nonumber
\Omega(u)                 &=& m(p)^2 u + m^2 (1-u) + p^2 u(1-u) 
\end{eqnarray}  
An approximate solution of eq.(2.12) may be obtained with the replacement of $\Omega(u)$
by $<\Omega> = m(p)^2/2 + p^2/6 $ in the $m=0$ limit, when eq.(2.12) reduces to
\begin{equation}\label{2.13}
z = x -\gamma + \ln x;  \quad x \equiv \mu^2/ <\Omega>; \quad z \equiv \pi /\alpha_s(\mu)
\end{equation}
This is a transcendental equation in $x$, whose approximate solution is
\begin{equation}\label{2.14}
x \approx z+\gamma - \ln(z+\gamma) + \frac{\ln(z+\gamma)}{z+\gamma} \equiv f(z)
\end{equation} 
Then an explicit solution for $m(p)$ is found from the last two equations as
\begin{equation}\label{2.15}
m(p)^2 = 2\mu^2 / f(z) - p^2/3
\end{equation}
A simpler solution which nevertheless incorporates the bulk (non-perturbative)
effects, is obtained by neglecting the perturbative propagator, in which case
eq.(2.14) reduces to $ z=x$ only, so that the mass function acquires the simpler form
\begin{equation}\label{2.16}
m(p)^2 = 2\mu^2 /z - p^2 /3 = m_q^2 - p^2 /3; \quad m_q^2 \equiv 2 \mu^2 \alpha_s(\mu)/ \pi 
\end{equation}
which is a slight improvement over the corresponding result of [11].

\section{Evaluation of 1-Loop Effective Potential $V_1$}

\setcounter{equation}{0}
\renewcommand{\theequation}{3.\arabic{equation}}

We shall now use the result (2.16) for $m(p)$ to evaluate  the effective potentials
$V_1$ given by eq.(2.6),  by the method of DR [18]. [The next section deals with the
corresponding two-loop potential $V_2$, eq.(2.7)]. 
Denoting the integrals of Eq. (2.6) by $V_{1i}$ $i= 1-4$, we first consider $V_{11}$
to illustrate the steps of the DR method [18]. Thus we write
\begin{equation}\label{3.1}
V_{11} = - \mu^{4-n} \int \frac{2d^n p}{(2\pi)^n} \ln(2/3 + m_q^2 / p^2)
\end{equation}
where we have as usual [11] supplied a compensating dimensional factor $\mu^{4-n}$ in front
and have substituted from eq.(2.16). Using eq.(2.11), we now get 
\begin{equation}\label{3.2}
V_{11} = \frac{2\pi^2\mu^4}{(2\pi)^4}[\frac{3 m_q^2}{2\mu^2}]^{n/2}\Gamma(-n/2)
\end{equation}     
where, in the $\pi$ factors in front, we have set $n=4$ [11], which may be 
regarded as coming under a `modified' minimal subtraction scheme [17]. Setting $n=4-\epsilon$
and subtracting the $\epsilon = 0$ pole contribution [11], gives finally
\begin{equation}\label{3.3}
V_{11} = + \frac{\pi^2 \mu^4}{(2\pi)^4} y^2 [ 3/2 - \gamma - \ln y ]; \quad y \equiv 3m_q^2 / 2\mu^2
\end{equation}
Proceeding in an exactly similar way, the other terms in (2.6) may be evaluated.Thus
\begin{equation}\label{3.4}
V_{12} = - \mu^{4-n} \int \frac{4d^n p}{(2\pi)^n} \frac{p^2}{m^2(p) +p^2}   
       = - \frac{6\pi^2 \mu^4}{(2\pi)^4} y^2 [ 1 - \gamma - \ln y ]; \quad y \equiv 3m_q^2 / 2\mu^2
\end{equation}
\begin{equation}\label{3.5}
V_{13} = - \mu^{4-n} \int Tr \frac{2 d^n p}{(2\pi)^4} \ln[1 + \frac{2\mu^2}{p^2+m^2}] 
       = + \mu^4 \frac{\pi^2}{(2\pi)^4}[ 3/2 - \gamma - \ln 2]
\end{equation}
\begin{equation}\label{3.6}
V_{14} = \mu^{4-n} \int \frac{2\pi^n}{(2\pi)^n} (1+ \frac{2\mu^2}{p^2 +m^2}) \Rightarrow ZERO
\end{equation} 
the last integral vanishing on taking $m=0$. This completes the evaluation of $V_1$. 

\section{Evaluation Of 2-Loop Effective Potential $V_2$ }

\setcounter{equation}{0}
\renewcommand{\theequation}{4.\arabic{equation}}

We now turn to the two-loop potential $V_2$ defined by (2.7) as a double 4D integral. 
A convenient strategy is first to integrate w.r.t. $d^4 p$ over the two fermionic
propagators so as to give rise to a gluon self-energy operator. The second integral
w.r.t. $d^4k$ then gives a vacuum self-energy graph by joining up the 2 gluon lines.
To organize the integral, we first define the gluon self-energy operator as
\begin{equation}\label{4.1}
\Pi_{\mu\nu}(k)= \mu^{(4-n)}\int \frac{g_s^2 F_1.F_2 d^n p}{2(2\pi)^n} Tr [S_F'(p)\gamma_\mu S_F'(p-k)\gamma_\nu]
\end{equation}
where $S_F'$ is given by (2.4). Then 
\begin{equation}\label{4.2}
V_2 = \mu^{4-n}\int \frac{d^n k}{(2\pi)^n} \Pi_{\mu\nu}(k) \Delta_{\mu\nu}(k)
\end{equation}
where $\Delta_{\mu\nu}$ is given by (2.5) in the Landau gauge. To evaluate (4.1), we 
substitute from (2.4) and (2.16), introduce the Feynman variable $ 0 \leq u \leq $ to combine
the two denominators, take the traces, give a translation $ p \rightarrow p+uk$
and drop the odd terms. Because of DR [18], one should expect gauge invariance to be
satisfied automatically, were it not for the approximate solution (2.16) which militates
against it. To meet this requirement we may still resort to the old-fashioned method [20,21]
of `gauge regularization' to extract the gauge invariant terms. The result of all these 
steps is the gauge invariant operator 
\begin{equation}\label{4.3}
\Pi_{\mu\nu}(k) = (k^2 \delta_{\mu\nu}- k_\mu k_\nu) \int_0^1 du u(1-u)  
      \frac{9\pi^2 g_s^2 F_1.F_2\Gamma(2-n/2) \mu^{4-n}}{(2\pi)^4[3m_q^2/2 + k^2 u(1-u)]^{(2-n/2)}} 
\end{equation}
where we have also carried out the dimensional integral over $d^n p$ using the formula (2.10). Next
we do DR [18] a la [11]. This gives
\begin{equation}\label{4.4}
\frac{\Gamma(2-n/2) \mu^{4-n}}{[3m_q^2/2 + k^2 u(1-u)]^{(2-n/2)}} \Rightarrow
 - [\gamma + \ln \frac{k^2 u(1-u) + 3 m_q^2/2}{\mu^2}]
\end{equation}
Substitution of (4.3-4) and (2.5) in (4.2) gives on simplification the $k$- integral for $V_2$:
\begin{eqnarray}\label{4.5}
V_2  &=& - 3g_s^2 F_1.F_2 \frac{9\pi^2}{(2\pi)^4}\int_0^1 du u(1-u) \int \frac{d^n k}{(2\pi)^4} \\
\nonumber
     & & \times \mu^{(4-n)}[\gamma + \ln \frac{k^2 u(1-u) + 3 m_q^2/2}{\mu^2}] [ 1 + \frac{2\mu^2}{k^2 + m^2}]
\end{eqnarray}
where the factor $3$ in front comes from the simplification of the $k_\mu$ factors 
in the Landau gauge. To organize the integral, note first that the $\gamma$ term does not
contribute in the $m=0$ limit. There are now two terms, $V_{21}$ and $V_{22}$, associated
with the non-perturbative ($\mu$-term) and perturbative (1-term)  contributions respectively. 
Both integrals may be carried out a la formula (2.11). The results are:
\begin{equation}\label{4.6}
V_{21} = -4 g_s^2 \frac{9\pi^4}{(2\pi)^8} (3 m_q^2 \mu^2) \int_0^1 du 
        [\frac{2\mu^2 u(1-u)}{3 m_q^2}]^{(2-n/2)} \frac{\Gamma(1-n/2)}{(n/2 - 1)};
\end{equation}
\begin{equation}\label{4.7}
V_{22} = -4 g_s^2 \frac{9\mu^4\pi^4}{(2\pi)^8} \int_0^1 du u(1-u) \Gamma (-n/2)  
[\frac{3 m_q^2}{2\mu^2 u(1-u)}]^{n/2}
\end{equation}  
We now need to do DR [18] on both these integrals just like in the pieces of $V_1$ above.
The case of $V_{21}$ which has a simple pole at $n=4$, is straightforward and gives
\begin{equation}\label{4.8}
V_{21} = -4 g_s^2 \frac{9\pi^4}{(2\pi)^8}(3 m_q^2 \mu^2) [-\gamma + \ln \frac{2\mu^2}{3 m_q^2}]
\end{equation}
where we have carried out an elementary integration over $u$ in the process. 
The other quantity $V_{22}$ is somewhat different in structure from the others since
the DR [18] can be effected only after the $u$-integration which leads to 
\begin{equation}\label{4.9}
V_{22} = -4 g_s^2 \frac{9\pi^4}{(2\pi)^8} (3 m_q^2 /2)^2 \frac{\Gamma^2(2-n/2) \Gamma(-n/2)}{\Gamma(4-n)} 
\end{equation}
This is a new feature which shows up as a $double pole$ in $4-n = \epsilon$, so that DR now
involves subtraction of $both$ the negative powers of $\epsilon$ before collecting
the finite terms when $\epsilon \rightarrow 0$.The steps are facilitated by the 
following expansions (for small $x$) [17]:
\begin{equation}\label{4.10}
\Gamma(x) = x^{-1}-\gamma + \frac{x}{2}[\gamma^2 + \pi^2 /6]; \quad
\Gamma^3(1+x) = 1- 3\gamma x + 3x^2 (\gamma^2 + \pi^2 /6)/2
\end{equation}        
The result is 
\begin{equation}\label{4.11}
V_{22}  = -4 g_s^2 \frac{9\pi^4}{(2\pi)^8} (3 m_q^2 /2)^2  
        [-(3\gamma) \ln (y) + \ln^2 (y)/2 -3\gamma +7 / 4 + \gamma^2 /2 -\pi^2 / 12]           
\end{equation}
where 
\begin{equation}\label{4.12}
y = \frac{3 m_q^2}{2\mu^2}; \quad g_s^2 = 4\pi \alpha_s(\mu) = 4\pi^2 y/3  
\end{equation}               
the last one coming from the relation (2.16), viz., $m_q^2 = 2\mu^2 \alpha_s(\mu)/\pi$. 
Substitution in (4.8) and (4.11) then gives   
\begin{eqnarray}\label{4.13}
V_{21} &=& 6C y^2 (\gamma + \ln y); \quad C= \frac{\pi^2 \mu^4}{(2\pi)^4}  \\  \nonumber
V_{22} &=& 3C y^3 [(\gamma - 3) \ln y + \ln^2(y)/2 -3\gamma +7/4 +\gamma^2/2 - \pi^2/ 12]
\end{eqnarray}
In the same notation we also record the expressions for $V_{1i}$ from Section 3 as
\begin{eqnarray}\label{4.14}
V_{11} &=& C y^2 ( 3/2 \gamma - \ln y); \quad V_{12} = - 6C y^2 (1- \gamma - \ln y) \\  \nonumber
V_{13} &=& C [ 3/2 - \ln 2 -\gamma]; V_{14} = ZERO
\end{eqnarray}        

\section{ Results and Discussion}

\setcounter{equation}{0}
\renewcommand{\theequation}{5.\arabic{equation}}

We are now in a position to use the results of (4.13-14) to determine the relation 
of the confining parameter $\mu$ with the QCD scale parameter $\Lambda_{qcd}$ by
demanding the minimality of the total effective potential $F(y) =V_1 + V_2$ regarded 
as a function of the ratio $y$, while holding $\mu$ fixed. Namely, $F'(y) =0$
which after factoring out the trivial solution $y=0$, simplifies to
\begin{equation}\label{5.1}
f(y) \equiv 2+22\gamma +22 \ln y + 3y[ -3+\gamma + (3\gamma-8)\ln y +1.5 \ln^2 y -1.911] =0
\end{equation}
This yields the result
\begin{equation}\label{5.2}
y \equiv \frac{3\alpha_s(\mu)}{\pi} = 0.475; \quad \alpha_s = \frac{2\pi}{9 \ln(\mu / \Lambda_{qcd})}  
\end{equation}
which provides the desired connection
\begin{equation}\label{5.3}
\Lambda_{qcd}= \mu (0.246) 
\end{equation}          
This result may be compared with the input values used in [11], viz., $\mu= 1 GeV$ and 
$\Lambda_{qcd}= 200 MeV$, taken from the spectroscopic data [14]. Thus the theoretical
value agrees with the empirical inputs to within about $20 \%$. To see the effect of 
including the 2-loop effects, the value obtained from minimising the the one-loop potential only, viz.,
$$ F_1(y) \equiv y^2 (3/2 -\gamma - \ln y) -6 y^2 (1-\gamma- \ln y) $$
yields $\ln y = 0.4 -\gamma$, leading to the estimate 
$$ \Lambda_{qcd} = 0.4512 \mu $$
which is more than double the input value [11,14]. Thus the inclusion of
the two-loop contribution is crucial for self-consistency in the determination.  

\subsection{Significance of $\mu$ Parameter}
 
One should now ask " what is the theoretical status of $\mu$ vis-a-vis
$\Lambda_{qcd}$" ? For while the latter is well-rooted in RG theory which is structured 
on pQCD, the introduction of the former in a more or less ad hoc manner demands a formal
placement within the QCD framework. A conservative view would be to regard $\mu$ as a 
sort of intermediate scale which controls the value of $\alpha_s$ in the strong-interaction
regime of confinement. Indeed its  modest value of $\sim 0.5$, eq.(5.2) corresponds 
precisely to such a regime, just as its (much smaller) values corresponding to the 
heavier masses of $W, Z$ bosons are more appropriate to the electroweak regime. For
a more formal basis to $\mu$ one needs to go back to a closer look at the connection (1.1)
between the original gluon field $A_\mu$ and a new one $B_\mu$, with an obvious convention
that in the hermitian conjugated relation the derivative acts from right to left. While the 
total content of the QCD Lagrangian remains unaltered, the latter can in principle be 
reformulated in terms of the fields $B_\mu$ and $B_\mu^\dag$ by writing $A_\mu$ as 
\begin{equation}\label{5.4}
2A_\mu = (1 + \sigma \partial_m)^{-1} B_\mu + B_\mu^\dag (1 - \sigma \partial_m)^{-1} 
\end{equation}
And although the total QCD content remains the same, the emphasis on a $B$-field
centred perturbative formalism clearly implies a more efficient incorporation of 
confinement effects than is usually possible in terms of the $A$-field, much like 
an improved convergence often achieved with a more efficient convergence parameter.  
Indeed the (more or less exact) derivation of the gluon propagator in Appendix A
should bring out the efficacy of the $B$-field. A more concrete analogy is perhaps 
to the ``dynamical perturbation theory'' of Pagels and Stokar [22], which effectively 
incorporates a good deal of QCD information in its vertex structure, so that its $loop$
 diagrams can afford to be largely free from criss-cross gluon lines. So far in this 
paper we have not exploited the fuller potential of the $B$-field, except for the 
derivation (in Appendix A) of the propagator (1.2) which already accounts for a bulk of 
non-perturbative (confinement) effects via the $B$-field description.  
And while in this paper, the exercise has been confined merely to a consistency check 
on the value of the $\mu$ parameter via the minimality of the effective action up to 
2-loop terms (to demonstrate that it is not a free parameter), the possibility of
a more systematic approach to non-perturbative QCD in terms of the $B_\mu$ fields,
with their associated Feynman diagrams etc, is clearly indicated.   
\par                 
	In conclusion, the present approach to confinement in QCD is only
one of many such attempts since the inception of QCD [1-7]. In recent times there
has been a spurt of like investigations (though not directly comparable with the 
physics of the present one) such as those dealing with the string structure of QCD, 
especially with a $vector$ type confinement [23,24], or the Seiberg-Witten theory of 
flux-tubes [25] 
which in turn has an obvious similarity with [23,24 ]. A cross section of other 
interesting approaches are  domain-like structures 
with chiral-symmetry breaking [26], Kugo-Ojima confinement criterion in Landau gauge 
QCD [27], and Chiral Lagrangian with confinement from the  QCD Lagrangian [28]. In the 
meantime the more time-honoured approaches like QCD-SR [12] and chiral perturbation 
theory [13] for the simulation of strong interaction effects which have already shown 
extensive evidence of flexibility in applications, are more amenable to comparison with
the present approach, as evidenced from the few results already found in [11]. On the 
basis of this limited comparison, we are optimistic that the
present approach offers a viable alternative to these non-perturbative methods, apart 
from an explicit incorporation of confinement in its basic formulation. 
\par
	I am grateful to Aalok Mishra and R.Ramanathan for useful comments on the
perspective of this paper.

\section*{Appendix: "Exact" Derivation Of Eq.(1.2) For $\Delta (k)$}

\setcounter{equation}{0}
\renewcommand{\theequation}{A.\arabic{equation}}

We give here a short derivation of the form (1.2) for the confining gluon propagator
through a simple process of iteration connecting the $A$ and $B$ fields, starting from 
eq.(1.1). The iteration consists in writing the solution of eq.(1.1) as a series
\begin{equation}\label{A.1}
A = A_0 + A_1 + A_2 + ... ; \quad B = B_0 + B_1 + B_2 +...  
\end{equation}
where $A_0$ is the lowest order (unperturbed) value of the $A$-field. The lowest
order $B$ field may be defined as
\begin{equation}\label{A.2}
B_0 = \frac{1}{1-\sigma \partial_m} A_0; \quad B_0^\dag = \frac{1}{1+\sigma \partial_m} A_0
\end{equation} 
For the next order in the two fields, substitute (A.2) in (1.1) once again to get
\begin{eqnarray}\label{A.3}
A_1  &=& \frac{1}{1-\sigma^2\partial_m^2}(A_0/2)\times two = \frac{1}{1-\sigma^2\partial_m^2}A_0 \nonumber \\ 
B_1  &=& \frac{1}{1-\sigma\partial_m}A_1; \quad B_1^\dag = \frac{1}{1+\sigma\partial_m}A_1  
\end{eqnarray}
And so on. The law is now clear. Thus in general
\begin{equation}\label{A.4}
A_n  = \frac{1}{1-\sigma^2\partial_m^2}A_{n-1}; \quad 
B_n  = \frac{1}{1-\sigma\partial_m}A_n; \quad B_n^\dag = \frac{1}{1+\sigma\partial_m}A_n
\end{equation}
Substituting these successive values in the series (A.1) gives a simple geometric series
which sums up to
\begin{equation}\label{A.5}
A = \frac{1}{1 - \frac{1}{1-\sigma^2\partial_m^2}}A_0 = A_0 - \frac{1}{\sigma^2\partial_m^2}A_0
\end{equation}
which is the requisite solution. To proceed further, we must evaluate the second term
on the right of (A.5) which amounts to a double integration w.r.t. $m$ on $A_0$. The
result of this integration is most succinctly expressed in terms of the propagator
whose exact form in momentum space may be shown as
\begin{equation}\label{A.6}
\Delta (k) = \lim_{m=0}[1 - \frac{1}{\sigma^2\partial_m^2}](k^2 + m^2)^{-1}
\end{equation}
which, $prima facie$, looks quite different from (1.2), but can be brought to this form
through a suitable choice of $\sigma$ which may be thought to `run' with $k^2$. Now
to evaluate the r.h.s. of (A.6), the result of two successive integrations gives
\begin{equation}\label{A.7}
\Delta (k) = \lim_{m=0} [\frac{1}{k^2 + m^2} + \frac{1}{2\sigma^2}\ln(\frac{k^2+m^2}{k^2+C})] 
\end{equation}     
where $C$ (like $\sigma$) is independent of $m$, and can again be chosen suitably.  Now a 
comparison of (A.7) with (1.2) suggests that we postulate 
$$ \sigma = k^2 / \mu_0 ; \quad C + k^2 = \Lambda^2 $$
where $\mu_0$ may have a further dependence on $k^2$. Now recall the definition of $\alpha_s$, viz.,
$$ \alpha_s = \frac{4\pi}{9 \ln(k^2/\Lambda^2)}$$
which suggests that the logarithmic factor would cancel out if we define $\mu_0^2 = \alpha_s \mu_1^2$
where $\mu_1^2$ is hopefully independent of $k^2$. Putting these things in (A.7) gives finally
\begin{equation}\label{A.8}
\Delta (k) = [ \frac{1}{k^2} + \frac{2\pi \mu_1^2}{9 k^4} ]
\end{equation}
which shows that (A.8) has exactly the form (1.2) if we have the correspondence (1.3) shown in text.

\end{document}